\documentstyle[12pt,epsfig,axodraw]{article}
\let\section=\subsection     \let\subsection=\subsubsection                

\newcommand{\M}{{\cal M}}

\newcommand{\GeV}{{\rm GeV}}

\newcommand{\fm}{{\rm fm}}
\renewcommand{\Im}{{\rm Im}}

\newcommand{\gsim}{$\raisebox{-0.8ex} {$\stackrel{\textstyle >}{\sim}$}$}

\begin{document}
\begin{center}
   {\large \bf QCD SUM RULES AND VECTOR MESONS }\\[2mm]
   {\large \bf IN NUCLEAR MATTER} \footnote{Work supported in part by GSI and BMBF}\\[5mm]
   F. Klingl and W. Weise  \\[5mm]
   {\small \it  Physik-Department\\ Technische Universit\"at
M\"unchen\\ Institut f\"ur Theoretische Physik\\ D-85747 Garching, Germany\\[8mm]}
\end{center}

\begin{abstract}\noindent
  Based on an effective Lagrangian which combines chiral SU(3) dynamics with
  vector meson dominance, we have developed a model for the s-wave vector
  meson-nucleon scattering amplitudes. We use this as an input for the low
  energy part of the current-current correlation function in nuclear matter.
  Its spectrum enters directly in the ``left hand side'' of QCD sum rules. For
  the isovector channel we find a significant enhancement of the in-medium
  spectral density below the $\rho$ resonance while the mass of $\rho$ meson
  itself decreases only slightly. The situation is different in the isoscalar
  channel. Here the mass and therefore the peak position of the $\omega$ meson
  decreases strongly while the width increases less drastically than in the
  $\rho$ meson case. For the $\phi$ meson we find almost no mass shift, the
  width of the peak broadens due to the imaginary part of the scattering
  length. We find a remarkable degree of consistency with the operator product
  expansion of QCD sum rules in all three channels.
\end{abstract}

\section{Introduction}
At present there is a lively discussion about the behaviour of vector mesons in
dense and hot hadronic matter. According to Brown and Rho \cite{1} the vector
meson masses should drop like the pion decay constant in baryonic matter
(BR-scaling). Other scaling laws \cite{1b} using a bag model find a mass
reduction of two third times that of the nucleon mass. Several analyses of the
$\rho$ and $\omega$ meson masses in matter using QCD sum rules \cite{2,3} seem
to confirm BR-scaling. On the other hand, model calculations of the density
dependent two pion self-energy of the rho meson in nuclear matter \cite{4,5}
suggest only marginal changes of the in-medium rho meson mass, but a strongly
increased decay width instead. Additional BR scaling seems to be needed in
order to match the hadronic models with the QCD sum rule analysis \cite{6}.
Phenomenologically, a dropping $\rho$ meson mass seems to help \cite{7,8}
understanding the enhanced dilepton yields seen at masses below the rho meson
resonance in the CERES and HELIOS experiments at CERN (see the updated review
by A. Drees at this workshop).

The QCD sum rule approach, at any practical level so far, has used only a
caricature of the true isovector spectrum, namely a parameterization in terms
of a delta function at the meson pole accompanied by a theta function type
continuum at higher masses. While such a parameterization is valid
in vacuum (as we shall confirm), this turns out not to be the case for the
in-medium spectrum. We will show that using a hadronic model to calculate the
vector meson-nucleon scattering lengths leads to correlation functions which
are consistent with the QCD sum rule analysis without need for additional BR
scaling.

In section 2 we review properties of the current
correlation function in vacuum and give a brief account of the QCD sum rule
approach. The current-current correlation functions in baryonic matter will be developed
in section 3. The comparison with QCD sum
rules at finite baryon density is made in section 4.

\section{Vector mesons and QCD sum rules in the vacuum}

Before we start looking into the in-medium properties of neutral vector mesons
it is useful to give a brief reminder of their vacuum properties. Since the vector mesons are
not stable they can only be seen as resonances in current-current (CC)
correlation function,
\begin{equation}
  \Pi_{\mu\nu}(q)=i\int d^4x \: e^{iq\cdot x}\langle
  0|{\cal T}j_{\mu}(x) j_{\nu}(0)|0\rangle  
 \label{2.1}
\end{equation} 
where ${\cal T}$ denotes the time-ordered product and $j_{\mu}$ is the
electromagnetic current. It can be decomposed as
\begin{equation}
  \label{2.3}
  j_{\mu}= j_\mu^\rho+j_\mu^\omega+j_\mu^\phi
\end{equation}
into vector currents specified by their quark
content:
\begin{eqnarray}
  j^\rho_{\mu}&=&\frac{1}{2}(\bar{u}\gamma_{\mu}u-\bar{d}\gamma_{\mu}d),\\
  j^\omega_{\mu}&=&\frac{1}{6}(\bar{u}\gamma_{\mu}u+\bar{d}\gamma_{\mu}d),\\
  j^\phi_{\mu}&=&-\frac{1}{3}(\bar{s}\gamma_{\mu}s).
\label{2.2}
\end{eqnarray}
Current conservation implies a transverse tensor structure
\begin{equation}
  \Pi_{\mu\nu}(q)=\left(g_{\mu\nu}-\frac{q_{\mu}q_{\nu}}{q^2}\right)\Pi(q^2),
\label{2.4}
\end{equation}
with the scalar CC correlation function
\begin{equation}
  \Pi (q^2)=\frac{1}{3} g^{\mu\nu}\Pi_{\mu\nu}(q).
\label{2.5}
\end{equation}
\begin{figure}[h]
\unitlength0.7071mm
\begin{picture}(200,180)
\put(0,0){\framebox(200,90){}}
\put(100,0){\line(0,1){90}}
\put(50,180){\line(1,0){100}}
\put(50,90){\line(0,1){90}}
\put(150,90){\line(0,1){90}}
\put(50,90){\makebox{\epsfig{file=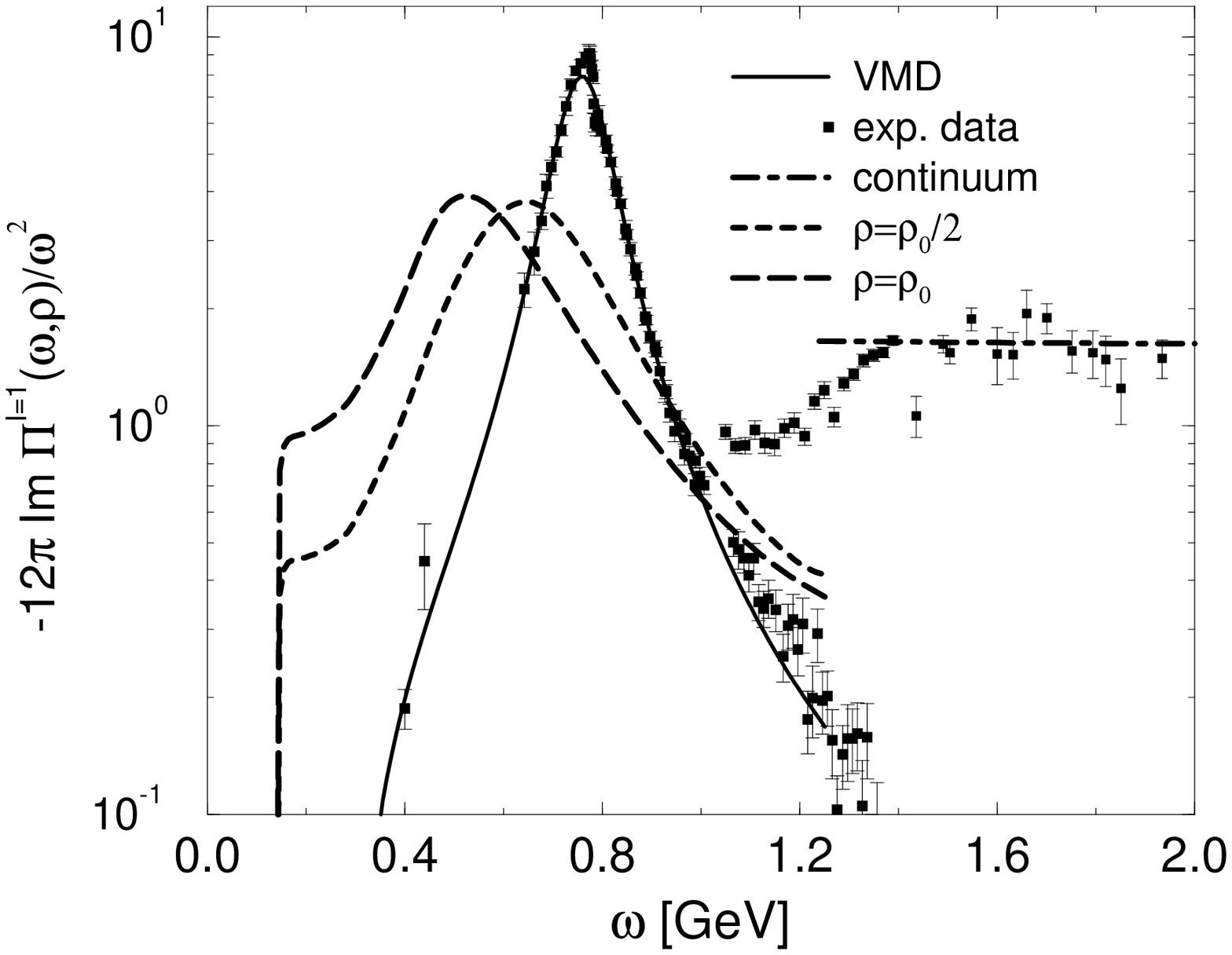,width=70mm}}}
\put(0,0){\makebox{\epsfig{file=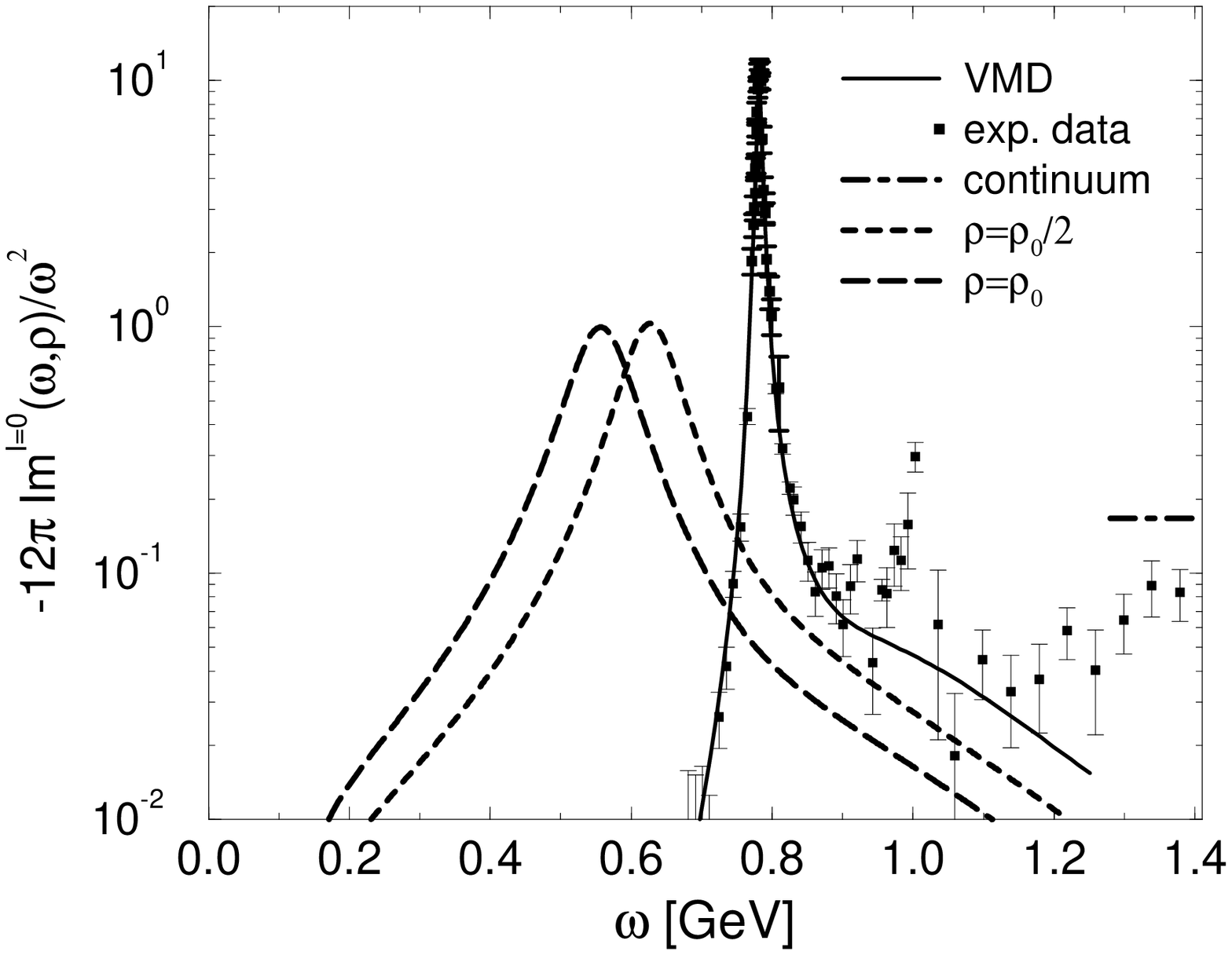,width=70mm}}}
\put(100,0){\makebox{\epsfig{file=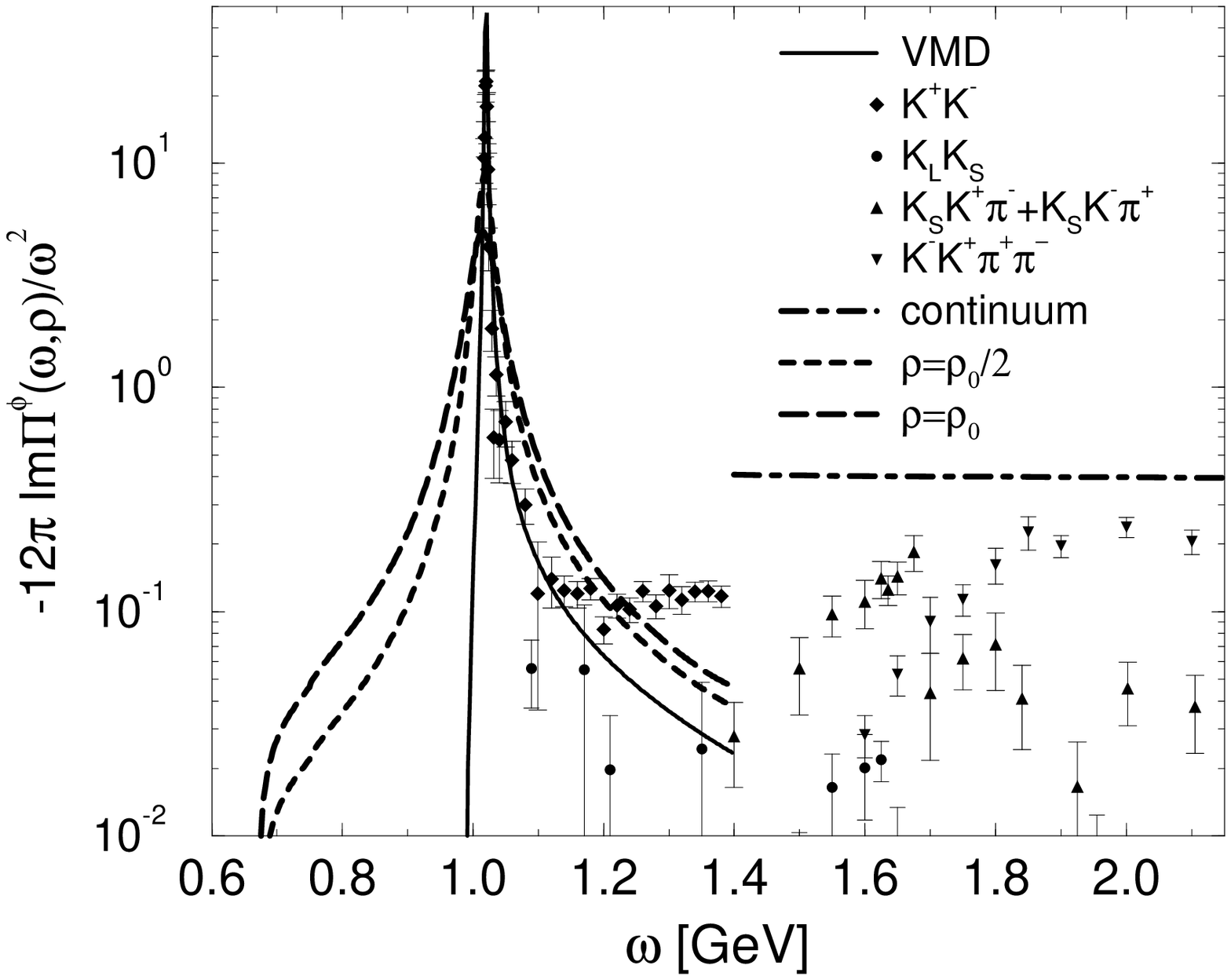,width=70mm}}}

\put(80,175){\makebox{a: $\rho$ meson}}
\put(30,85){\makebox{b: $\omega$ meson}}
\put(130,85){\makebox{c: $\phi$ meson}}
\end{picture}
\caption{Calculated spectra of current-current correlation functions. The solid
lines show the vacuum spectra in the $\rho$, $\omega$ and $\phi$ channels
normalized such that they can be compared directly with the corresponding
$e^+e^- \to$ hadrons data. The dashed lines show the spectral functions in
nuclear matter at densities $\rho_0/2$ and $\rho_0=0.17 \; \fm^{-3}$, as
discussed in section 3.}
\vspace*{-1cm}
\end{figure}

The imaginary part of the correlation function is proportional to the cross
section for $e^+ e^- \to$ hadrons: 
\begin{equation}
 R(s)=\frac{\sigma^{I=1}(e^+ e^-\to \rm hadrons )}{\sigma (e^+
  e^-\to \mu^+ \mu^-)}=-\frac{12 \pi}{s} \Im \Pi(s)
\label{2.6}
\end{equation}
where $\sigma (e^+ e^-\to \mu^+ \mu^-)= 4 \pi \alpha^2/3 s$ with $\alpha=e^2/4
\pi=1/137$. The vector mesons can be distinguished by looking at different
hadronic channels with corresponding flavour (isospin) quantum numbers.
For example, G-parity demands that the isovector current describing the $\rho$
meson can only decay into even numbers of pions (see data in fig.1a). Similarly
the annihilation of $e^+e^-$ into odd numbers of pions determines the isoscalar
current describing the $\omega$ meson.  This can be seen in figure 1b. Isospin
violating processes are small but visible as $\rho \omega$ mixing corrections
to the pion formfactor. The situation for the $\phi$ meson is more involved.
The OZI rule forbids decays into pions, however a strong violation of about
five percent is seen in the three-pion channel. (Some experimental data near
the $\phi$ resonance have been left out in fig. 1b to emphasize the
$\omega$-contribution given by the solid line). The $\phi$ meson decays mainly
into OZI allowed channels such as $K^+K^-$ or $K_LK_S$ (see fig. 1c). However
these channels also exist for the $\rho$ and $\omega$ mesons.  The measured
cross section of the annihilation into kaons includes an interference between
all three vector mesons. (For more details see ref \cite{12}).  Nevertheless
the $\phi$ meson still dominates the data in these channels.

Vector meson dominance (VMD) leads to the solid lines in figs.1a, b and c. In
comparison with the experimental data for $e^+ e^-$ annihilation into hadronic
channels, we clearly see that the low energy region of the CC correlation
function is very well described by VMD. The extended VMD model \cite{12} gives:
\begin{equation}
  \label{2.7}
  \Im \Pi (q^2)= \sum_V \frac{\Im \Pi_V^{\rm vac} (q^2)}{g_V^2} \left| F(q^2) \right|^2,
\end{equation}
where 
\begin{equation}
  \label{2.8}
   F(q^2) = \frac{(1-a_V) q^2-\stackrel{ \rm o }{m}_V^2}{q^2-\stackrel{ \rm o
   }{m}_V^2-\Pi_V^{\rm vac}(q^2)}\, .
\end{equation}
Here $\stackrel{ \rm o }{m}_V$ are the bare vector meson masses, $\Pi_V^{\rm
  vac}$ the vacuum self energies and $g_V$ the strong SU(3) couplings of the
vector mesons. The parameter $a_V$ is free, but usually close to one. It is
identical to unity in case of ``complete'' VMD in which all of the hadronic electromagnetic
interaction is transmitted through vector mesons. For the rho meson channel $F(q^2)$ is
simply given by the pion form factor
\begin{equation}
  \label{2.9}
   F_\pi (q^2) = \frac{\left(1- \frac{g}{\stackrel{ \rm o }{g}_\rho}\right)q^2-\stackrel{ \rm o }{m}_\rho^2}{q^2-\stackrel{ \rm o }{m}_\rho^2-\Pi_\rho(q^2)}\, ,
\end{equation}
where $g$ ($\stackrel{ \rm o }{g}_\rho$) is the coupling of the $\rho$ meson to
the pion (photon) and $\Pi_\rho$ is the $\rho$ meson self energy which is
dominated by the two-pion loop \cite{12}.

In the high energy region the measured correlation
function approaches the asymptotic plateau predicted by QCD:
\begin{equation}
  R (s)=d \left( 1+\frac{\alpha_S}{\pi} \right) \Theta (s- s_0)
\label{2.10}
\end{equation}
for large $s$, where $d_\rho=3/2$, $d_\omega=1/6$ and $d_\phi=1/3$. 

The basic idea of QCD sum rules is to compare two ways of determining $\Pi(q^2)$ in
the region of large spacelike  $Q^2=-q^2 \gg 1 \; \GeV$. One way is to use a twice
subtracted dispersion relation for each one of the channels $V=\rho$,
$\omega$, $\phi$:
\begin{equation}
  \Pi(q^2)=\Pi (0)+c \, q^2+\frac{q^4}{\pi}\int ds\frac{\Im\Pi (s)}{s^2(s-q^2
  -i\epsilon)}, 
\label{2.12}
\end{equation}
where the vanishing photon mass in the vacuum requires $\Pi(0)=0$. The other
one is to calculate the correlation function using the operator product
expansion (OPE):
\begin{eqnarray}
  \label{2.13}
  \frac{12 \pi}{Q^2} \Pi(q^2=-Q^2)& =&
  \frac{d}{ \pi} \left[-(1+\alpha_S(Q^2)/\pi) 
  \ln{\left(\frac{Q^2}{\mu^2}\right)} \right. \\ \nonumber && +\left. \frac{c_1}{Q^2}+\frac{c_2}{Q^4}+\frac{c_3}{Q^6}+... \right].
\end{eqnarray}
Here the coefficients $c_{1,2,3}$ incorporate the non-perturbative parts coming
from the condensates such as the gluon condensate in $c_2$ or the four-quark
condensate in $c_3$. The commonly used values \cite{14} for the isoscalar and
isovector channel are $c^\rho_2=c_2^\omega=0.04 \, \GeV^4$ and $
c^\rho_3=c_3^\omega=-0.07\, \GeV^6 $. The coefficient $c_1$ is proportional to
the squared quark mass; due to the small mass of the up and down quark we can
safely neglect those contributions and set $c_1^{\rho,\omega}=0$. The situation
is different in case of the $\phi$ meson. Because of the large strange quark
mass, $c_2$ changes and $c_1$ is no longer negligible. Here we take
$c^\phi_1=-0.07\, \GeV^2$, $c^\phi_2=0.17 \, \GeV^4$ and $ c^\phi_3=-0.07 \,
\GeV^6$ \cite{14}.

A Borel transform is used in order to improve the convergence of the OPE
series. Comparing eq.(\ref{2.12}) and (\ref{2.13}) after Borel transformation we end up with
\begin{eqnarray}
  \label{2.14} 
  \frac{12 \pi^2
  \Pi(0)}{d \M^2}+\frac{1}{
  d \M^2} \int_0^{\infty}ds R (s)
  \exp{\left[-\frac{s}{\M^2}\right]}= \nonumber \\
 \hspace{2cm}  (1+\alpha_S(\M^2)/\pi)+\frac{c_1}{\M^2}+\frac{c_2}{\M^4}+\frac{c_3}{2 \M^6},
\end{eqnarray}
where the Borel mass parameter should be chosen in the range $\M \gsim 1 \,
\GeV$ in order to ensure convergence of the OPE side of eq.(\ref{2.14}).  The
``left side'' comes from the dispersion relation of eq.(\ref{2.12}) and $R$
represents the ratio (\ref{2.6}), but now specified for each individual flavour
channel with $V=\rho,\omega,\phi$. The ``right side'' is determined by the OPE.
In fig. 3a, b and c we show the comparison between the ``left side'' (solid
line) and the ``right side'' (dashed line) for the various (vacuum) channels.
The consistency between ``left'' and ``right'' sides is evidently quite
satisfactory. Only for a Borel mass below $0.8 \, \GeV$ higher order terms in
the OPE become important and the convergence fails. Keeping this success of QCD
sum rules in mind we now want to explore whether this still holds in the
medium.

\section{Current correlation functions in baryonic matter}

The CC-correlation function in medium at temperature $T=0$ is defined as
\begin{eqnarray}
  \label{3.1}
  \Pi_{\mu\nu}(\omega ,\vec{q};\rho)\: \: = \:\:  i\int\limits^{+\infty}_{-\infty}dt\int
  d^3 x  \: e^{i\omega t-i\vec{q}\cdot\vec{x}} \hspace{2cm}\\ \nonumber \hspace{3.cm}
*\,  \left<\rho\left|{\cal T}j_{\mu}(t,\vec{x})j_{\nu}(0)
  \right|\rho\right>,
\end{eqnarray}
where we have replaced the vacuum by $\big| \rho \rangle$, the ground state of infinite, isotropic and isospin
symmetric nuclear matter with density $\rho$. We assume matter as a whole to be
at rest. This specifies the
Lorentz frame that we will use in the following. For $\vec{q}=0$, the case where the vector meson is at rest, only the
transverse tensor structure survives and $\Pi^{00}=\Pi^{0j}=\Pi^{i0}=0$. We write
\begin{equation}
  \label{3.2}
  \Pi_{ij} (\omega ,\vec{q}=0;\rho)= -\delta_{ij}  \Pi (\omega,\vec{q}=0;\rho).
\end{equation}
As a first step we apply the low density theorem to both the correlation
function and the self energy of the vector mesons. We have
\begin{equation}
\label{3.3}
  \Pi (\omega ,\vec{q}=0;\rho) = \Pi_{\rm vac}(\omega^2)-\rho \, T (
  \omega)+... \, , 
\end{equation}
and
\begin{equation}
\label{3.4}
  \Pi_V (\omega ,\vec{q}=0;\rho)= \Pi_V^{\rm vac}(\omega^2)-\rho \, T_{VN} (
  \omega)+... \, ,
\end{equation}
where $T(\omega)$ is the Compton amplitude and $T_{VN}$ is the vector
meson-nucleon scattering amplitude, both taken at $\vec{q}=0$. On the other
hand using eq.(\ref{3.4}) we can extend eq.(\ref{2.7}) to the in-medium
correlation function in terms of VMD and write \cite{11}
\begin{eqnarray}
  \label{3.5}
   g_V^2 \Im \Pi(\omega,\vec{q}=0;\rho)\: \: =\: \:\Im \left(\Pi^{\rm vac}_V(\omega^2)-\rho
   T_{V N}(\omega)\right)\\ \nonumber \hspace{2.0cm} * \;\left|\frac{(1-a_V) \,\omega^2-\stackrel{ \rm o
  }{m}_V^2}{\omega^2-\stackrel{ \rm o }{m}_V^2-\left(\Pi^{\rm
  vac}_V(\omega^2) -\rho \, T_{VN}(\omega)\right)}\right|^2.
\end{eqnarray}
Comparing the terms linear in density of eqs.(\ref{3.3}) and (\ref{3.5}) we
find that the Compton amplitude translated into VMD language becomes:
\begin{equation}
  \label{3.6}
  T(\omega)= g_V^2 \, F(\omega^2) \, T_{VN}(\omega)  \, F(\omega^2).
\end{equation}
In order to determine the in-medium correlation function we are left with
the task to calculate the vector meson-nucleon scattering amplitudes. We use an
effective Lagrangian which combines chiral SU(3) dynamics with VMD. This
Lagrangian has been applied successfully in the vacuum \cite{12} and has been
extended to incorporate meson-baryon interactions. For the baryons we include
nucleons, hyperons and $\Delta$'s. 
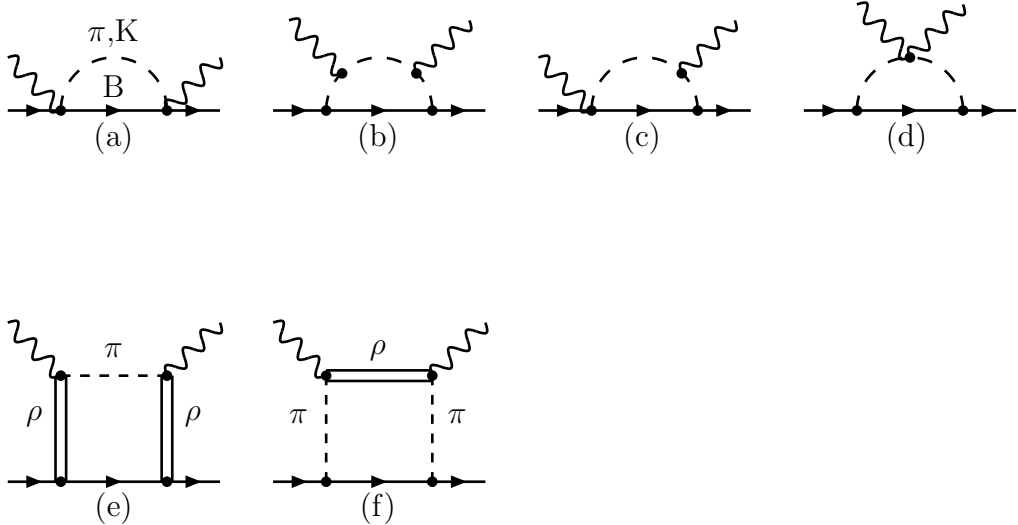
\begin{figure}[h]
\vspace*{-3cm}
\begin{center}
\begin{picture}(400,300)(0,0)
\SetWidth{1}    
\ArrowLine(10,155)(30,155)
\ArrowLine(30,155)(70,155)
\ArrowLine(70,155)(90,155)
\DashCArc(50,155)(20,0,180){6}
\Photon(10,175)(30,155){3}{3}
\Photon(90,175)(70,155){3}{3}
\Text(50,145)[]{(a)}
\Text(50,185)[]{$\pi$,K}
\Text(50,165)[]{B}
\Vertex(30,155){2}
\Vertex(70,155){2}

\ArrowLine(110,155)(130,155)
\ArrowLine(130,155)(170,155)
\ArrowLine(170,155)(190,155)
\DashCArc(150,155)(20,0,180){6}
\Photon(116,189)(136,169){3}{3}
\Photon(164,169)(184,189){3}{3}
\Text(150,145)[]{(b)}
\Vertex(130,155){2}
\Vertex(170,155){2}
\Vertex(136,169){2}
\Vertex(164,169){2}

\ArrowLine(210,155)(230,155)
\ArrowLine(230,155)(270,155)
\ArrowLine(270,155)(290,155)
\DashCArc(250,155)(20,0,180){6}
\Photon(210,175)(230,155){3}{3}
\Photon(264,169)(284,189){3}{3}
\Text(250,145)[]{(c)}
\Vertex(230,155){2}
\Vertex(270,155){2}
\Vertex(264,169){2}

\ArrowLine(310,155)(330,155)
\ArrowLine(330,155)(370,155)
\ArrowLine(370,155)(390,155)
\DashCArc(350,155)(20,0,180){6}
\Photon(330,195)(350,175){3}{3}
\Photon(350,175)(370,195){3}{3}
\Text(350,145)[]{(d)}
\Vertex(330,155){2}
\Vertex(370,155){2}
\Vertex(350,175){2}

\ArrowLine(10,15)(30,15)
\ArrowLine(30,15)(70,15)
\ArrowLine(70,15)(90,15)
\Line(28,15)(28,55)
\Line(32,15)(32,55)
\Line(68,15)(68,55)
\Line(72,15)(72,55)
\DashLine(30,55)(70,55){4}
\Photon(10,75)(30,55){3}{3}
\Photon(70,55)(90,75){3}{3}
\Text(50,5)[]{(e)}
\Text(20,40)[]{$\rho$}
\Text(80,40)[]{$\rho$}
\Text(50,65)[]{$\pi$}
\Vertex(30,15){2}
\Vertex(70,15){2}
\Vertex(30,55){2}
\Vertex(70,55){2}

\ArrowLine(110,15)(130,15)
\ArrowLine(130,15)(170,15)
\ArrowLine(170,15)(190,15)
\Line(130,53)(170,53)
\Line(130,57)(170,57)
\DashLine(130,15)(130,55){4}
\DashLine(170,15)(170,55){4}
\Photon(110,75)(130,55){3}{3}
\Photon(170,55)(190,75){3}{3}
\Text(150,5)[]{(f)}
\Text(120,40)[]{$\pi$}
\Text(180,40)[]{$\pi$}
\Text(150,65)[]{$\rho$}
\Vertex(130,15){2}
\Vertex(170,15){2}
\Vertex(130,55){2}
\Vertex(170,55){2}
\end{picture}
\end{center}
\caption{Dominant diagrams for the Compton amplitude $T(\omega)$ and the vector
meson-nucleon amplitudes $T_{VN}(\omega)$. In order to get from $T$ to $T_{VN}$,
replace the (wavy) photon line by the relevant vector mesons. Diagrams (e) and
(f) operate, in particular, in the $\omega$ meson channel.}
\end{figure}
The most important processes contributing to the scattering amplitudes are
shown in fig.2 a and b for the $\rho$, $\omega$ and $\phi$ meson respectively.
For the $\rho$ ($\phi$) meson we draw the diagrams which survive in the limit
of large baryon mass (fig. 2a-d). The last one (fig. 2d) only contributes to
the real part of the $\rho N$ ($\phi N$) scattering amplitude. While for the
$\rho$-meson the $\pi N$ and $\pi \Delta$ loops govern the scattering
amplitude, $K\Sigma$ and $K\Lambda$ loops dominate for the $\phi$ meson. For
the $\omega N $ scattering amplitude we only show the dominant contributions to
the imaginary part (figs. 3e, f). We evaluate the imaginary parts of the
amplitudes by cutting the diagrams in all possible ways. The real parts are
then determined by using a once subtracted dispersion relation, with the
subtraction constant fixed by the Thomson limit. Evaluating those diagrams and
using eq.(\ref{3.5}) we plot the spectra of the correlation functions for
various densities as shown by the dashed curves in fig. 2a, b and c. We observe
important differences between the various channels. The $\rho$ meson mass
decreases only slightly while the width increases very strongly. This causes
the peak to shift downwards and broaden. We also see strong threshold
contributions starting at the pion mass. On the other hand the mass of the
$\omega$ meson decreases significantly. Its width also increases but not as
strongly as for the $\rho$ meson. For the $\phi$ meson there is almost no
change in the peak position while the width increases.

\section{Comparison with QCD sum rules}
We now proceed to test the consistency of these in-medium correlation functions
with the OPE of QCD sum rules. Performing an OPE of the density dependent
correlation functions \cite{4} the coefficients of eq.(\ref{2.13}) become
density dependent.  We use the values
$c_2^{\rho,\omega}=0.04+0.018(\rho/\rho_0) \, \GeV^4$ and $
c^{\rho,\omega}_3=-0.07+0.036 (\rho/\rho_0) \GeV^6$ proposed by Hatsuda et al.
\cite{2} for the $\rho$ and $\omega$ meson. They neglected higher twist
condensates which are hardly known but might be important. For the four-quark
condensates we assume as in ref.\cite{3} that ground state saturation holds to
the same extent as in the vacuum.  For the $\phi$ meson the coefficient $c_1$
remains unchanged and we take the values $c_2^{\phi}=0.17+0.01 (\rho/\rho_0) \,
\GeV^4$ and $c^\phi_3=-0.07+0.006 (\rho/\rho_0) \GeV^6$ as suggested by Asakawa
and Ko \cite{13}. In figure 3 we have plotted the Borel transformed OPE at
normal nuclear matter (long dashed lines) and compared with the ``left side''
(dashed dotted line) as given by the dispersion relation representation of our
calculated in-medium correlation function.

The consistency at normal nuclear matter is not as excellent as in the
vacuum but still very impressive.  We therefore conclude that QCD sum rules
and our hadronic model of the in-medium current-current correlation
function are mutually compatible. We also point out that assuming a simple pole ansatz for the spectrum at finite densities can lead
to erroneous interpretation of the in-medium masses. This is very obvious in the isovector
channel. While our model gives only a small change of the rho mass a simple
pole ansatz leads to a reduction of $m_\rho$ by more than 10 percent \cite{2,3} at $\rho=\rho_0$.
\begin{figure}[h]
\unitlength0.7071mm
\begin{picture}(200,180)
\put(0,0){\framebox(200,90){}}
\put(100,0){\line(0,1){90}}
\put(50,180){\line(1,0){100}}
\put(50,90){\line(0,1){90}}
\put(150,90){\line(0,1){90}}
\put(50,90){\makebox{\epsfig{file=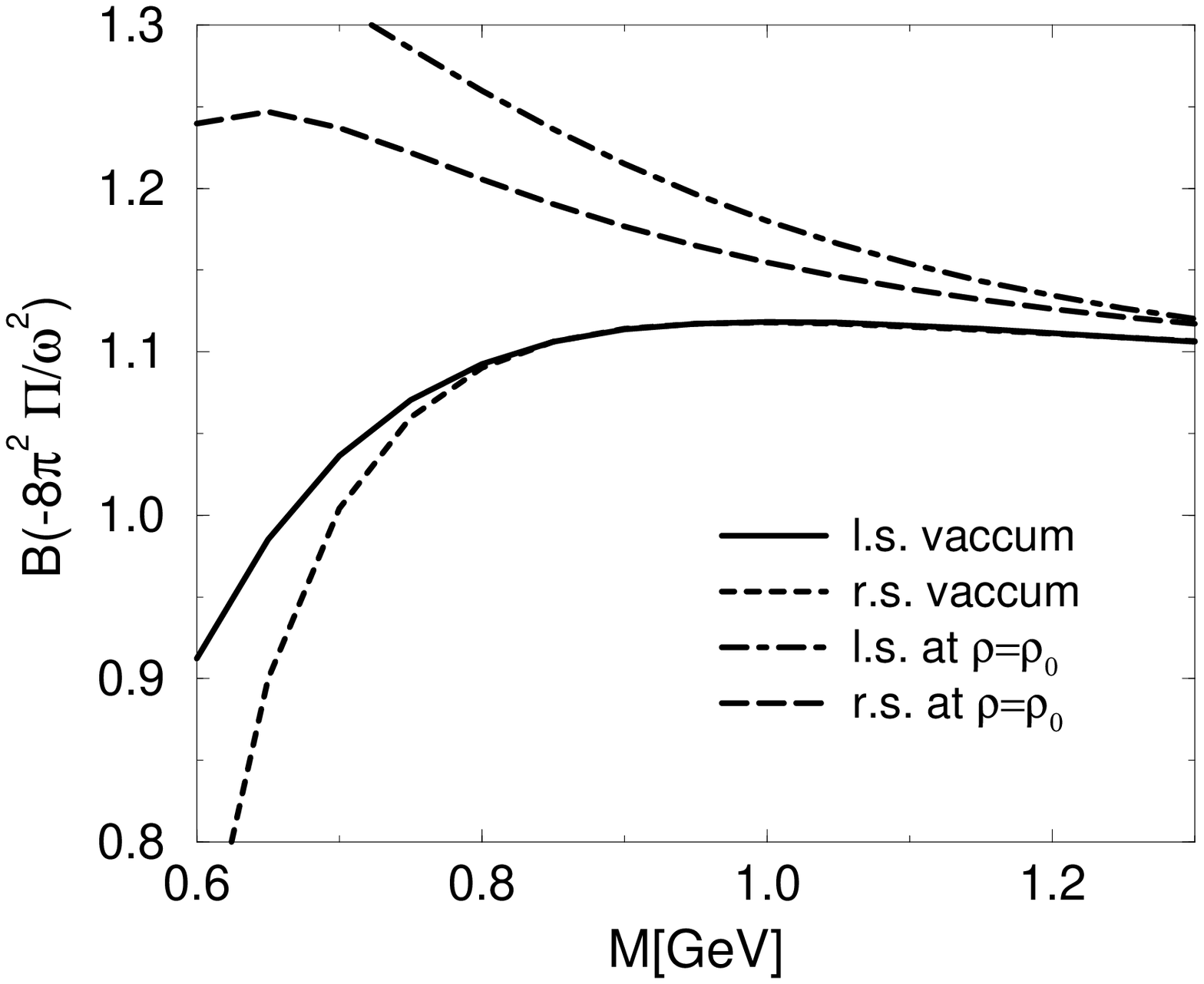,width=70mm}}}
\put(0,0){\makebox{\epsfig{file=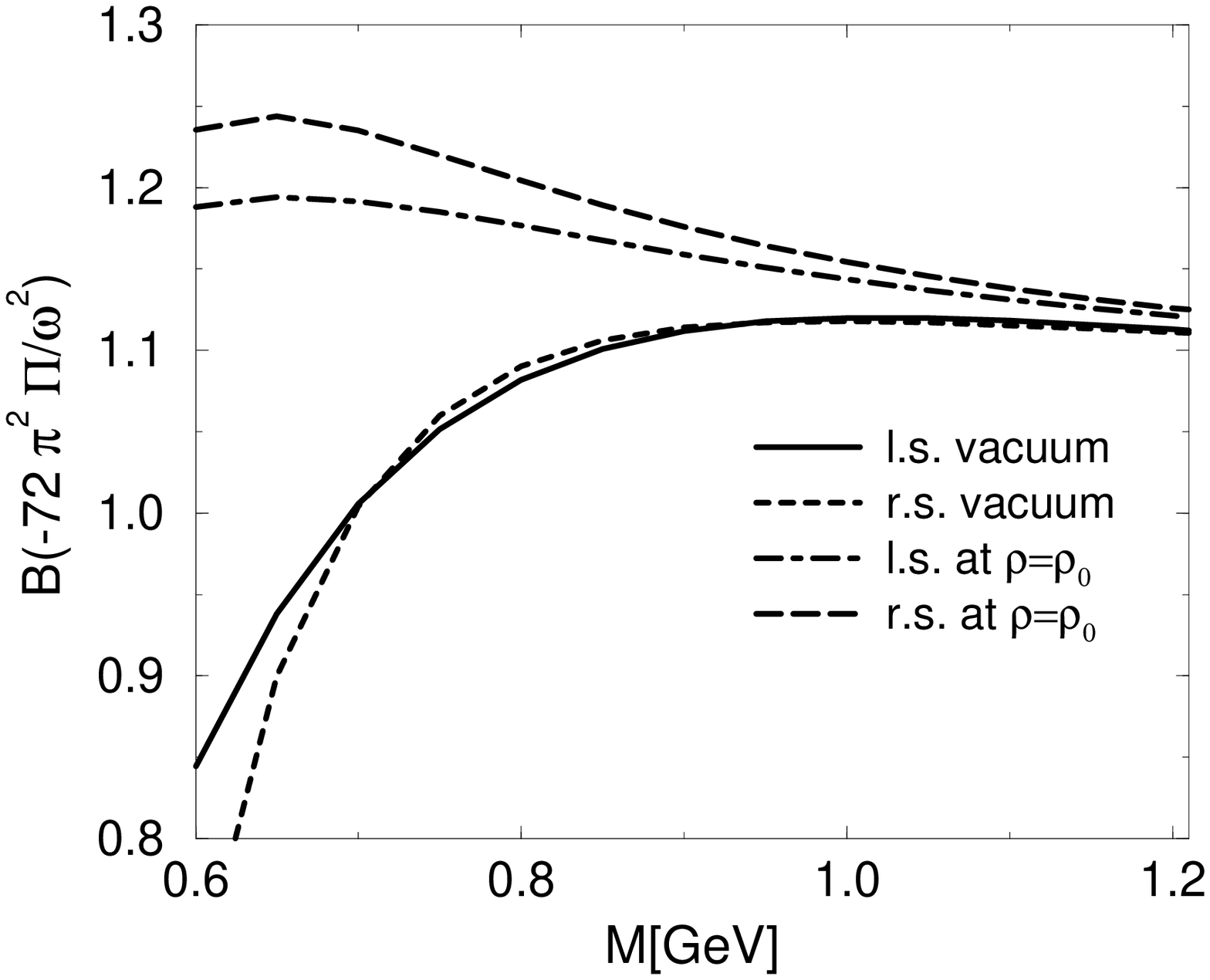,width=70mm}}}
\put(100,0){\makebox{\epsfig{file=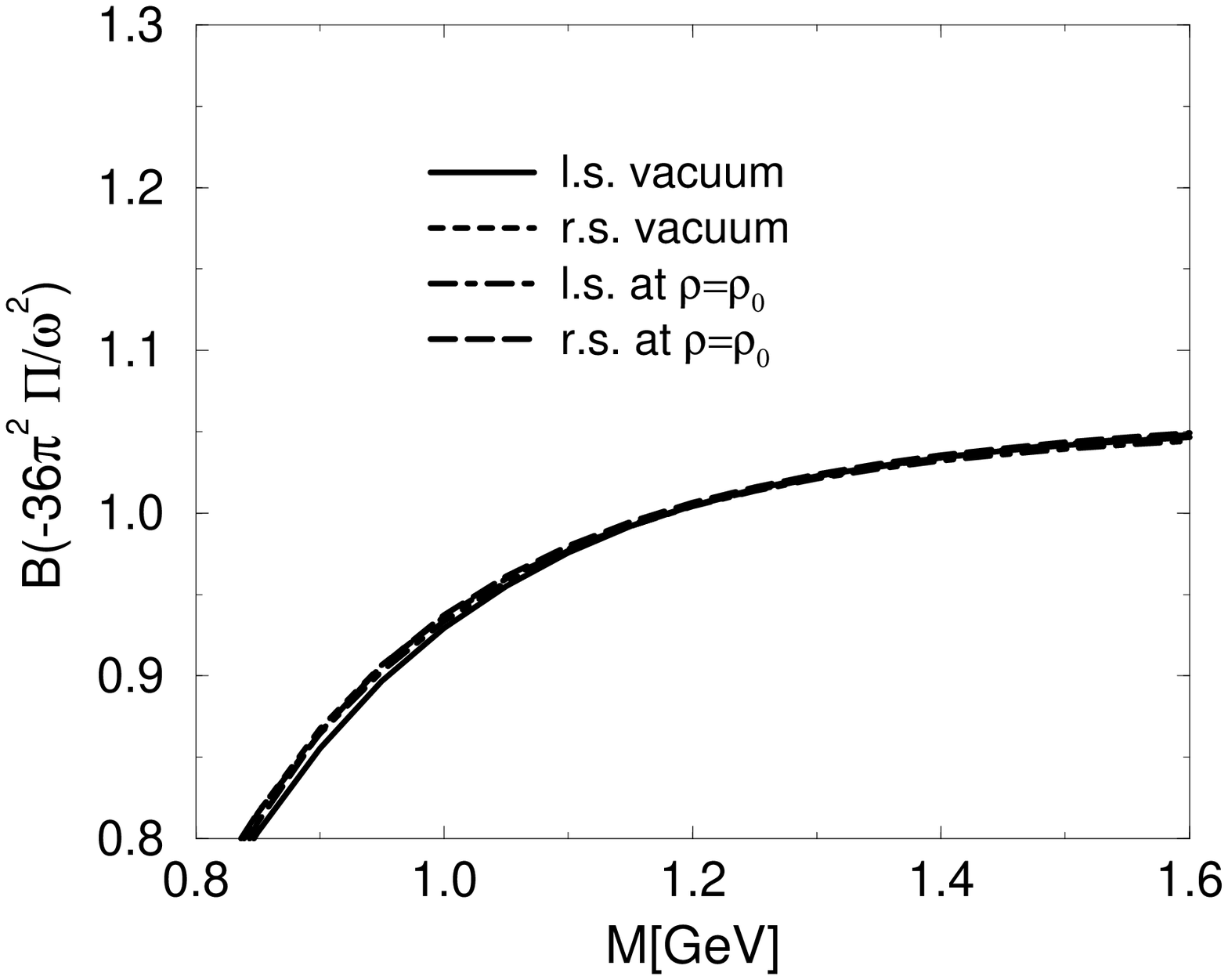,width=70mm}}}

\put(80,175){\makebox{a: $\rho$ meson}}
\put(30,85){\makebox{b: $\omega$ meson}}
\put(130,85){\makebox{c: $\phi$ meson}}
\end{picture}
\caption{Comparison of the ``left'' and ``right'' side of the QCD sum rules
(15) in the
vacuum (solid and dashed line) and at
normal nuclear density (dot dashed and long dashed line) as a function of the
Borel mass parameter { \protect \cal M}.}
\vspace*{-2cm}
\end{figure}


\begin{thebibliography}{99}
\itemsep=0cm
\bibitem{1}G.E. Brown and M. Rho, Phys. Rev. Lett. {\bf{66}} (1991)
2720.
\bibitem{1b} K. Saito, K. Tsushima and A.W. Thomas, Nucl Phys. {\bf A 609} (1996) 339.
\bibitem{2}T. Hatsuda and S.H. Lee, Phys. Rev. {\bf{C 46}} (1992) R34.
\bibitem{3}X. Jin and D.B. Leinweber, Phys. Rev. {\bf{C 52}} (1995) 3344.
\bibitem{4} M. Herrmann, B.L. Friman and W. N\"orenberg, Nucl. Phys. {\bf A
560} (1993) 411.
\bibitem{5} G. Chanfray and P. Schuck, Nucl. Phys. {\bf A
555} (1993) 329.
\bibitem{6}M. Asakawa and C. M. Ko, Phys. Rev. {\bf C 48} (1993) 526.
\bibitem{7}G.Q. Li, C.M. Ko and G.E. Brown,
Phys. Rev. Lett. {\bf 75} (1995) 4007.
\bibitem{8}W. Cassing, W. Ehehalt and C.M. Ko, Phys. Lett. {\bf{B 363}}
(1995) 35.
\bibitem{11} F. Klingl and W. Weise, Nucl. Phys. {\bf A 606} (1996) 329;
F. Klingl, N. Kaiser and W. Weise, preprint (1997).
\bibitem{12} F. Klingl, N. Kaiser and W. Weise, Z. Phys. {\bf A 356} (1996) 193.
\bibitem{13} M. Asakawa and C. M. Ko, Nucl. Phys. {\bf A 570} (1994) 732.
\bibitem{14} M.A. Shifman, A.I. Vainshtein and
V.I. Zakharov, Nucl. Phys. {\bf B 147} (1979) 385.
\end{thebibliography}
\end{document}